\documentclass[fleqn,twoside]{article}
\usepackage[headings]{espcrc2}

\readRCS
$Id: espcrc2.tex,v 1.2 2004/02/24 11:22:11 spepping Exp $
\usepackage{graphicx}
\usepackage[figuresright]{rotating}

\newcommand{\AmS}{{\protect\the\textfont2
  A\kern-.1667em\lower.5ex\hbox{M}\kern-.125emS}}

\title{The ALICE detector and trigger strategy for diffractive and
       electromagnetic processes}

\author{R. Schicker,\address[RS]{Physikalisches Institut,
        Philosophenweg 12, 69120 Heidelberg }%
        \thanks{This work is supported in part by German BMBF under
        project 06HD197D} on behalf of the ALICE Collaboration  }
       
\runtitle{2-column format camera-ready paper in \LaTeX}
\runauthor{S. Pepping}

\begin{document}

\begin{abstract}
The ALICE detector at the Large Hadron Collider (LHC) consists of a
central barrel, a muon spectrometer, zero degree calorimeters and
additional detectors which are used for trigger purposes and for event 
classification. The main detector systems of relevance for measuring 
diffractive and electromagnetic processes are described.   
The trigger strategy for such measurements is outlined. The physics
potential of studying diffractive and electromagnetic processes at the
LHC is presented by discussing possible signatures of the Odderon.      

\vspace{1pc}
\end{abstract}

% typeset front matter (including abstract)
\maketitle

\section{The ALICE Experiment}
The ALICE experiment is presently being built and commissioned at the
Large Hadron Collider (LHC)\cite{Alice1,Alice2}. The ALICE experiment 
consists of a central barrel covering the pseudorapidity 
range $-0.9 < \eta < 0.9$ and a muon spectrometer in the 
range $-4.0<\eta<-2.4$. Additional detectors for trigger purposes and 
for event classification exist such that the 
range $ -4.0 < \eta < 5.0 $ is covered. The ALICE physics program
foresees data taking in pp  and PbPb collisions at 
luminosities up to $\cal{L}$ = $5\times 10^{30}cm^{-2}s^{-1}$ and 
$\cal{L}$ = $10^{27}cm^{-2}s^{-1}$, respectively. 
An asymmetric system pPb will be measured at a luminosity of
$\cal{L}$ = $10^{29}cm^{-2}s^{-1}$.

\section{The ALICE Central Barrel}

The detectors in the ALICE central barrel track and identify 
hadrons, electrons and photons in the pseudorapidity range 
$ -0.9 < \eta < 0.9$. The magnetic field strength 
of 0.5 T  allows the measurement of tracks from very low transverse 
momenta  of about 100 MeV/c to fairly high values of about 100 GeV/c. 
The tracking detectors are designed to reconstruct secondary vertices 
resulting from decays of hyperons, D and B mesons. The granularity of
the central barrel detectors is chosen such that particle tracking and
identification can be achieved in a high multiplicity environment of
up to 8000 particles per unit of rapidity. The main detector systems
for these tasks are the Inner Tracking System, the Time Projection
Chamber, the Transition Radiation Detector and the Time of Flight
array. These systems cover the full azimuthal angle within the 
pseudorapidity range $ -0.9 < \eta < 0.9$ and are described below. 
Additional detectors with partial coverage 
of the central barrel are a PHOton Spectrometer (PHOS), an
electromagnetic calorimeter (EMCAL) and  a High-Momentum Particle 
Identification Detector (HMPID). 

\subsection{The Inner Tracking System}

The Inner Tracking System (ITS) consists of six cylindrical layers of
silicon detectors at radii from 4 cm to 44 cm. The minimum radius
is determined by the beam pipe dimensions whereas the maximum radius 
chosen is determined by the necessity of efficient track matching with
the outer detectors in the central barrel. 
The innermost layer extends over the range 
$ -2 < \eta < 2 $ such that there is continous overlap with 
event classification detectors outside of the central barrel.
Due to the high particle density of up to 80 particles/cm$^{2}$ and 
in order to achieve the required tracking resolution, pixel detectors 
have been chosen for the first two layers. Silicon drift detectors 
are located in the  middle two layers whereas double sided silicon
strip detectors are in the outer two layers. 
   
\subsection{The Time Projection Chamber}

The Time Projection Chamber (TPC) is the main tracking detector in the 
central barrel. The inner and outer radii of the active volume are 
84.5 cm and 246.6 cm, respectively. The full radial track length is 
measured in the pseudorapidity range $ -0.9 < \eta < 0.9$ whereas
tracks with at least one third of nominal radial length  are covered in the 
pseudorapidity range $ -1.5 < \eta < 1.5$. Particle identification
is achieved by measuring the specific ionization loss.    
The chosen geometry results in a drift time of about 90 $\mu$s. This
long drift time is the factor limiting the proton-proton luminosity
to the value mentioned above.   
   
\subsection{The Transition Radiation Detector}

The principal goal of the Transition Radiation Detector (TRD) is to
provide electron identification in the momentum range larger than 
1 GeV/c. In this range, the electron identification by energy
loss in the TPC is no longer sufficient. Since the TRD is a
fast tracker, the TRD information can be used for an efficient trigger 
on high transverse  momentum electrons. In addition, the position 
information from the TRD system improves the tracking performance of 
the central barrel. 

\subsection{The Time of Flight Detector}

The Time-Of-Flight (TOF) array is located at a radial distance 
from 3.7 m to 4.0 m. 
The TOF information is used for particle identification in the
range 0.2 GeV/c $ < p_{T} < $ 2.5 GeV/c. 
For this detector, the Multi-gap Resistive-Plate (MRPC) technology was 
chosen. A  strip with an active area of 120x7.4 cm$^{2}$ consists 
of pads of 3.5 cm length and 2.5 cm width.  
  
\subsection{The Central Barrel Performance}

\subsubsection{Tracking Resolution}

The ITS, TPC and TRD detectors described above are the main tracking 
detectors in the central barrel. With the information from these 
detectors, particles with momenta as low as 100 MeV/c can be tracked.

\begin{figure}[htb]
\includegraphics*[scale=0.36]{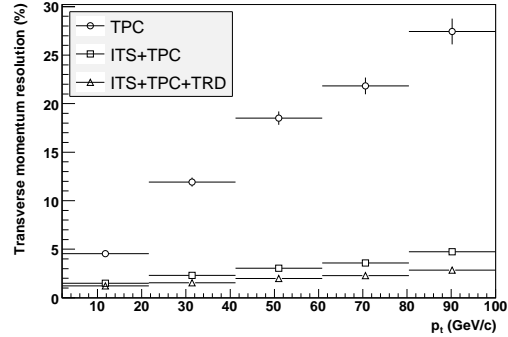}
%\vspace{9pt}
\caption{Central barrel tracking resolution}
\label{fig:trl_perf}
\end{figure}

Fig.\ref{fig:trl_perf} shows the transverse momentum resolution
as expected from simulations. The TPC alone achieves a resolution of
approximately 3\% at a transverse momentum of $p_{T}$ = 10 GeV/c.
Adding the information from ITS and TRD on the inner and outer side,
respectively, improves the resolution considerably due to the increased
leverage. The combined transverse momentum  resolution from the ITS, TPC and
TRD detector is expected to be about 3\% at a transverse momentum
of $p_{T}$ = 100 GeV/c.

\subsubsection{Particle identification}

Particle identification is achieved in the central barrel by 
different methods. First, the specific energy loss is
measured by the TPC, the TRD and
the strip and drift detectors of the ITS. 

\begin{figure}[htb]
\includegraphics*[scale=0.40]{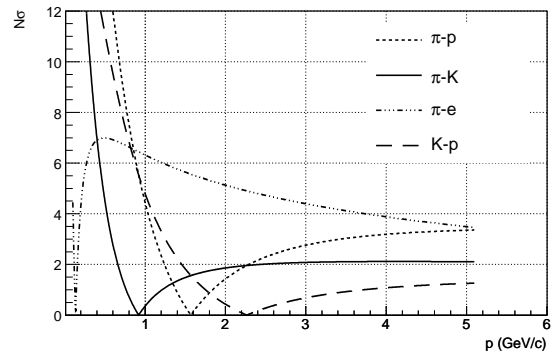}
%\vspace{9pt}
\caption{Particle identification by dE/dx measurement}
\label{fig:dedx_pid}
\end{figure}

Fig.\ref{fig:dedx_pid} shows the combined particle identification 
capability  by dE/dx measurement as a function of momentum.  The 
separation of different particle species is shown in units of the 
resolution of the dE/dx measurement. 

\begin{figure}[htb]
\includegraphics*[scale=0.54]{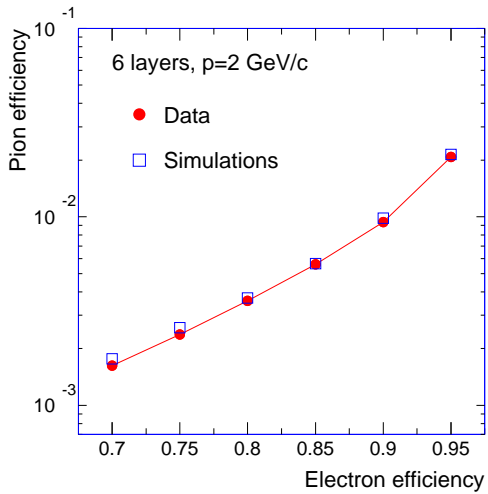}
%\vspace{9pt}
\caption{Electron-pion separation in the TRD}
\label{fig:eff_elec_pion}
\end{figure}

The electron-pion separation at high momenta is significantly improved 
by the information of the TRD system. Fig.\ref{fig:eff_elec_pion}
shows the pion efficiency in the TRD as a function of
the electron efficiency. Here, pion efficiency indicates that a pion
is misidentified as an electron. The expected TRD performance for a
full stack of six layers  is shown by the squares and compared 
to test beam data represented by the circles.

\section{The ALICE Zero Degree Neutron Calorimeter}

The Zero Degree Neutron Calorimeters (ZDC) are placed on both sides of the 
interaction point at a distance of 116 m\cite{ZDC}. The
ZDC information can be used to select different diffractive
topologies. Events of the type $pp \rightarrow ppX$ do not deposit energy
in these calorimeters, events $pp \rightarrow pN^{*}X$ will have
energy in one of the calorimeters whereas events 
$pp \rightarrow N^{*}N^{*}X$ will have energy deposited in both calorimeters. 
Here, X denotes a centrally produced diffractive state from which the 
diffractive L0 trigger is derived as described below.

\section{The ALICE diffractive gap trigger}

Additional detectors for event classification and trigger purposes
are located on both sides of the ALICE central barrel. First, an array
of scintillator detectors (V0) is placed on both sides of the 
central barrel. These arrays are labeled V0A and V0C on the 
two sides, respectively. Each of these arrays covers a pseudorapidity
interval of about two units with a fourfold segmentation of half a 
unit. The azimuthal coverage is divided into eight 
segments of 45$^{0}$ degrees hence each array is composed  of 32
individual counters.  
Second, a Forward Multiplicity Detector (FMD) is located on both sides 
of the central barrel. The pseudorapidity coverage of this detector
is $-3.4 < \eta < -1.7$ and $1.7 < \eta < 5.1$, respectively.

\begin{figure}[htb]
\includegraphics*[scale=0.34]{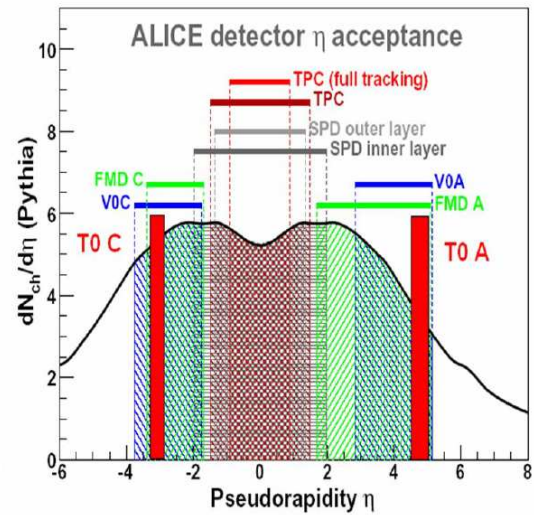}
%\vspace{9pt}
\caption{Pseudorapidity coverage of trigger detectors and of detectors
in central barrel}
\label{fig:acc}
\end{figure}

Fig.\ref{fig:acc} shows the pseudorapidity coverage of the detector
systems described above. The geometry of the ALICE central barrel 
in conjunction with the additional detectors V0 and FMD is well suited
for the definition of a rapidity gap trigger. The ALICE trigger system
is designed as a multi-level scheme with L0,L1 and L2 levels and a 
high-level trigger (HLT). 
A central trigger processor (CTP) collects the information from the
different detector systems and distributes the L0,L1 and L2 decisions 
back to the detectors. For L0 decision, the trigger information has 
to arrive at the CTP 850 ns after the interaction time. 
A rapidity gap trigger can be defined by the requirement of signals 
coming from the central barrel detectors while V0 and FMD not showing 
any activity. Such a scheme requires a trigger signal from within 
the central barrel for L0 decision. The pixel detector of the ITS 
system is suited for delivering such a signal\cite{pixel}. 

The TRD detector system needs special consideration in the definition
of a rapidity gap trigger. The TRD readout electronic is partly put in sleep
mode after readout of an event in order to reduce power
consumption. The TRD detector hence has its own pretrigger system
which generates a wake up signal prior to an L0 decision from CTP.
This pretrigger system has access to the information of the V0
detectors and generates wake up calls based thereupon. 
The V0 signal is, however, absent in a diffractive trigger.
The pixel information described above is late for a
diffractive TRD wake up call. Such a diffractive TRD wake up call can
be generated based upon the information of the TOF detector.
The information of the TOF array is forwarded to the TRD pretrigger
system where multiplicity conditions and topological constraints
are defined. In addition, the information of the V0
detectors is available at this level hence the rapidity gap width 
can be defined. The resulting L0 signal is fast enough to reach
the CTP well before the time limit for L0 decision.      

The high level trigger HLT has access to the information of all the 
detectors shown in Fig.\ref{fig:acc} and will hence be able to select
events with rapidity gaps in the range $-4 < \eta < -1$ and 
$1 < \eta < 5$. These gaps extend over seven units of pseudorapidity
and are hence expected to suppress minimum bias inelastic events
by many orders of magnitude.   

In addition to the scheme described above, the ALICE diffractive L0 
trigger signal can be generated from the Neutron ZDC if no central state 
is present in the reaction. A L0 signal from ZDC does, however,
not meet the 850 ns L0 time constraint described above. A L0 trigger
from ZDC is therefore only possible during special data taking runs 
for  which the 850 ns time limit is extended by  about 150 ns. 
The possibility of such data taking is currently under discussion.     

\section{ALICE diffractive and electromagnetic physics}

The tracking capabilities at very low transverse momenta in
conjunction with the excellent particle identification make ALICE an 
unique facility at LHC to pursue a long term physics program of
diffractive and electromagnetic physics. The low luminosity of ALICE
as compared to the other LHC experiments restricts the ALICE physics 
program to reactions with cross section at a level of a few nb per unit 
of rapidity.

\begin{figure}[htb]
\includegraphics*[scale=0.28]{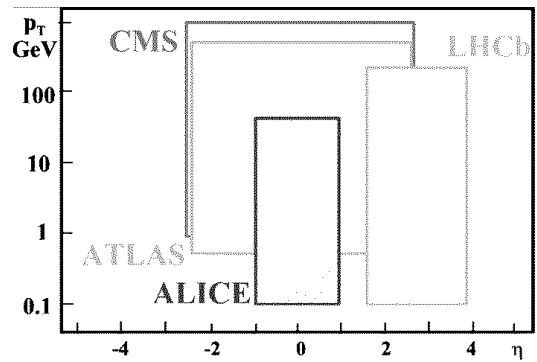}
%\vspace{9pt}
\caption{Rapidity and transverse momentum acceptance of the LHC experiments}
\label{fig:acc_all}
\end{figure}

Fig.\ref{fig:acc_all} shows the transverse momentum acceptance of the 
four main LHC experiments. Not shown in this
figure is the acceptance of the TOTEM experiment which has a
physics program of measurements of total cross section, elastic
scattering and soft diffraction\cite{TOTEM}. The acceptance of the
TOTEM telescopes is in the range of $ 3.1 <  | \eta |  < 4.7$ and 
$5.3 < | \eta | <6.5$. The combined data taking of TOTEM and CMS  
represents the largest rapidity interval covered at the LHC. 
The CMS transverse momentum acceptance of about 1 GeV/c shown
in Fig.\ref{fig:acc_all} represents a nominal value. The CMS analysis 
framework foresees the reconstruction of a few selected data samples to
values as low as 0.2 GeV/c\cite{CMS}.   
 
\section{Signatures of the Odderon}

The Odderon was first postulated in 1973 and is represented  
by color-singlet exchange with negative C-parity\cite{nicolescu}. Due to its
negative C-parity, Odderon exchange can lead to differences between
particle-particle and particle-antiparticle scattering. In QCD, 
the Odderon can be a three-gluon object in a symmetric color state.
Due to the third gluon involved in the exchange, a suppression by the 
coupling $\alpha_s$ is expected as compared to the two gluon Pomeron
exchange. However, finding experimental signatures of the Odderon 
exchange has so far turned out to be extremely difficult\cite{ewerz}.
A continued non-observation of Odderon signatures 
would put considerable doubt on the formulation of high energy
scattering by gluon exchange\cite{pomeron_qcd}. The best evidence 
so far for Odderon 
exchange was established as a difference between the differential
cross sections for elastic $pp$ and $p\bar{p}$ scattering 
at $\sqrt{s}$ = 53 GeV at the CERN ISR. The $pp$ cross section
displays a dip at t = -1.3 GeV$^2$ whereas the $p\bar{p}$ cross
section levels off. Such a behaviour is typical for negative
C-exchange and cannot be due to mesonic Reggeons only.   

\subsection{Signatures of Odderon Cross Sections}

Signatures of Odderon exchanges can be looked for in exclusive 
reactions where the Odderon (besides the photon) is the only possible
exchange. Diffractively produced C-even states such as pseudoscalar
or tensor mesons can result from photon-photon, photon-Odderon and
Odderon-Odderon exchange. Any excess measured beyond the well
understood photon-photon contribution indicates an Odderon
contribution.

Diffractively produced C-odd states such as vector mesons 
$\phi, J/\psi, \Upsilon$ can result from photon-Pomeron or 
Odderon-Pomeron exchange. Any excess beyond the photon contribution
would be indication of an Odderon exchange. 

\begin{figure}[htb]
\includegraphics*[scale=0.96]{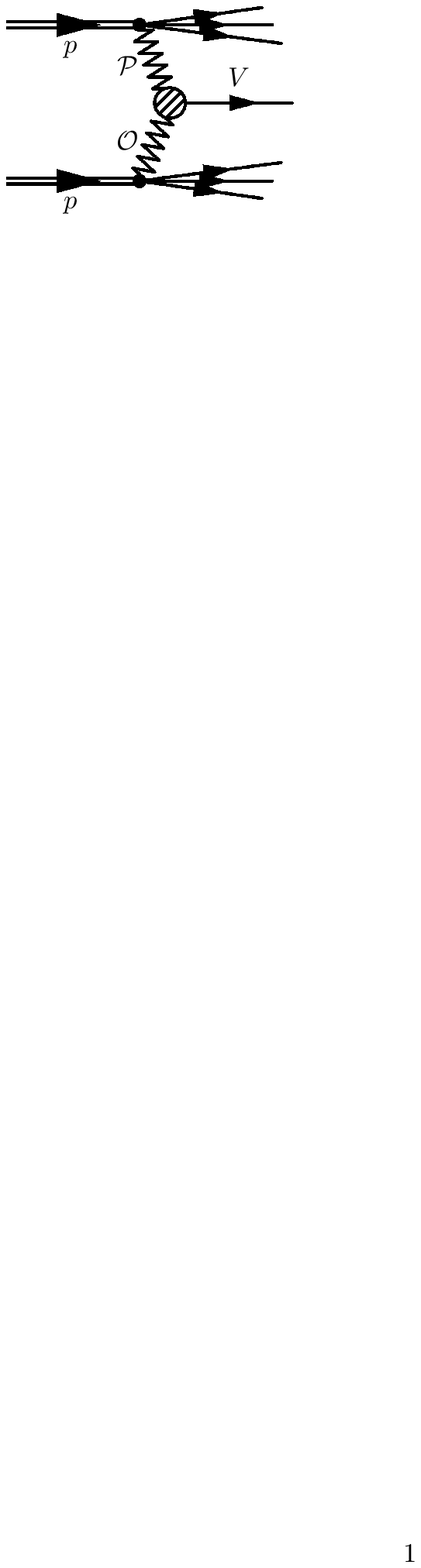}
\includegraphics*[scale=0.80]{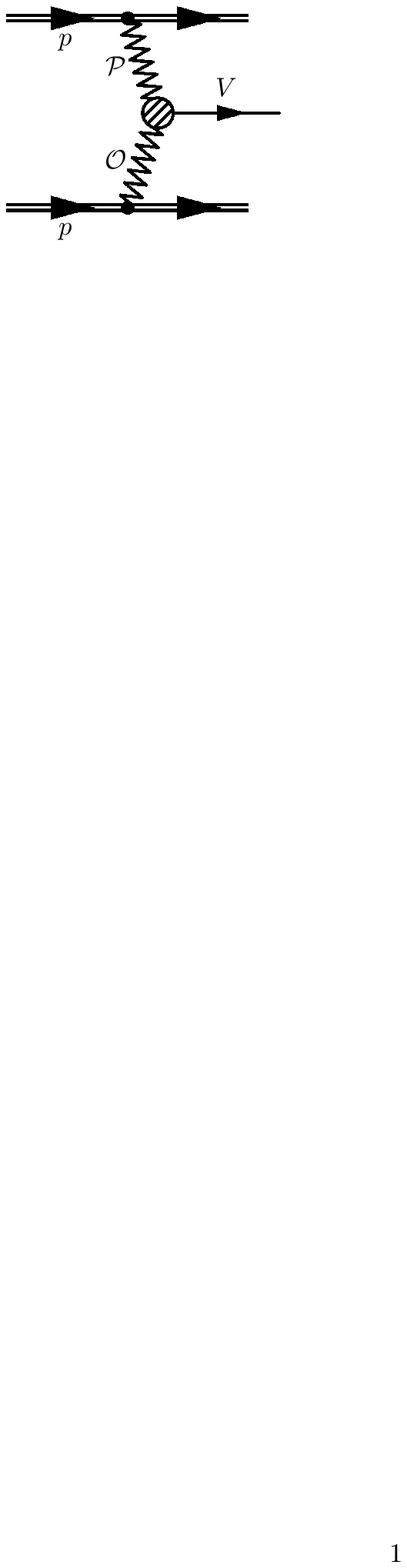}
%\vspace{9pt}
\caption{Vector Meson production by Odderon-Pomeron fusion}
\label{fig:odderon}
\end{figure}

Fig.\ref{fig:odderon} shows the Feynman diagram for vector meson
production by Pomeron-Odderon fusion with breakup of the protons on
the left and without breakup on the right. The two different reaction 
channels can be identified by the information of the ZDC. To each of
the two diagrams in Fig.\ref{fig:odderon} exists a diagram in which
the Odderon is replaced by a photon.  

\begin{figure}[htb]
\includegraphics*[scale=0.7]{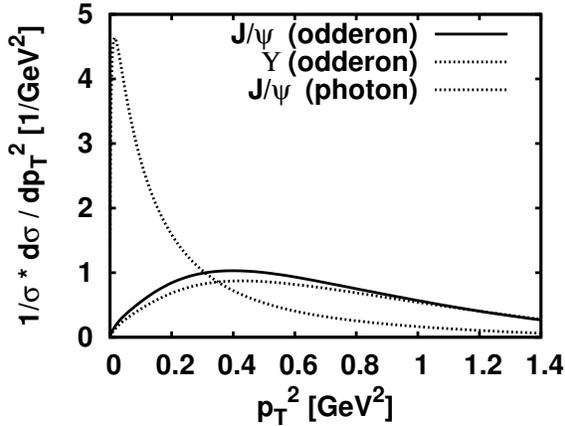}
%\vspace{9pt}
\caption{The J/$\psi$ transverse momentum distribution for
the photon and Odderon contributions}
\label{fig:odderon_pt}
\end{figure}

Cross sections for diffractively produced $J/\psi$ in pp
collisions at LHC energies were first estimated by Sch\"{a}fer et 
al\cite{schaefer}. More refined calculations by Bzdak et al result in 
a t-integrated photon contribution of $\frac{d\sigma}{dy}\mid_{y=0} \;
\sim$ 15 nb and a t-integrated Odderon contribution of   
$\frac{d\sigma}{dy}\mid_{y=0} \; \sim$ 1 nb\cite{bzdak}. 
These two numbers carry large uncertainties, the upper and lower limit 
of these numbers vary by about an order of magnitude. This cross section
is, however, at a level where in 10$^6$ s of data taking in ALICE the 
$J/\psi$ can be measured in its e$^+$e$^-$ decay channel at a level 
of 4\% statistical uncertainty. Due to the different t-dependence, the 
two  contributions result in different $p_T$ distributions of the $J/\psi$. 
The photon and Odderon contributions are shown in Fig.\ref{fig:odderon_pt} 
by the dotted and solid lines, respectively. A careful transverse 
momentum analysis of the $J/\psi$ might therefore allow to disentangle
the Odderon contribution. 

\subsection{Signatures of Odderon Interference Effects}

If the diffractively produced final state is not an eigenstate of
C-parity, then interference effects between photon-Pomeron and
photon-Odderon amplitudes can be analyzed. 

\begin{figure}[htb]
\begin{center}
\includegraphics*[scale=1.0]{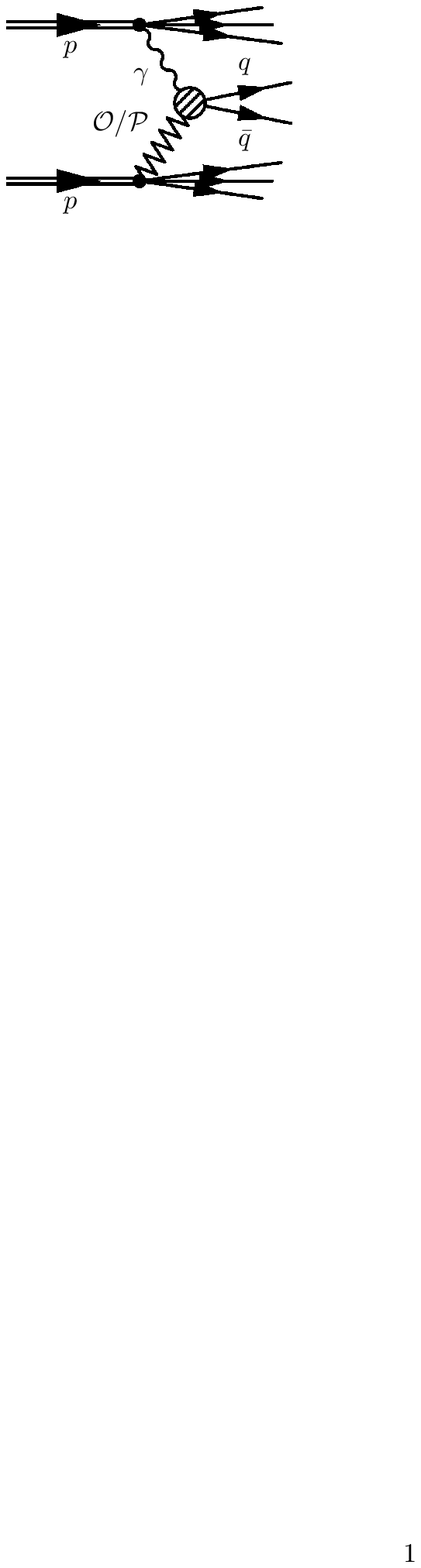}
%\vspace{9pt}
\caption{photon-Pomeron and photon-Odderon amplitudes}
\label{fig:odderon_inter}
\end{center}
\end{figure}

Fig.\ref{fig:odderon_inter} shows the photon-Pomeron and the
photon-Odderon amplitudes for $q\bar{q}$ production.  

A study of open charm diffractive photoproduction estimates the asymmetry in 
fractional energy to be on the order of 15\%\cite{brodsky}. The 
forward-backward charge asymmetry in diffractive production of pion
pairs is calculated to be on the order of  10\% for pair masses in the 
range $1\: GeV/c^{2} < m_{\pi+\pi-} < 1.3\: GeV/c^{2}$\cite{haegler,ginzburg}. 

\vspace{1cm}

{\bf Acknowledgments}

I  thank Otto Nachtmann and Carlo Ewerz for 
illuminating discussions and Leszek Motyka for preparing and
communicating Figure \ref{fig:odderon_pt}.


\begin{thebibliography}{9}

\bibitem{Alice1} F. Carminati et al, ALICE Collaboration, 2004, J.Phys. G:
  Nucl. Part. Phys. {\bf 30} 1517

\bibitem{Alice2} B. Alessandro et al, ALICE Collaboration, 2006, J.Phys. G:
  Nucl. Part. Phys. {\bf 32} 1295

\bibitem{ZDC}R. Arnaldi et al, Nucl. Instr. and Meth. A {\bf 564}
  (2006) 235

\bibitem{pixel} The ALICE collaboration, ALICE experiment at the CERN
  LHC, accepted for publication in JINST

\bibitem{TOTEM} K. Eggert, TOTEM a different LHC experiment, CERN
  colloquium, feb 21,2008 

\bibitem{CMS} D. d'Enterria et al, Addendum CMS technical design report, 
J. Phys. G34:2307-2455, 2007

\bibitem{nicolescu}L. Lukaszuk, B. Nicolescu, Lett. Nuovo Cim. {\bf 8}
  (1973) 406

\bibitem{ewerz}C. Ewerz, Proceedings XII Rencontres de Blois (2005) 377

\bibitem{pomeron_qcd} S. Donnachie, G. Dosch, P.V. Landshoff, O. Nachtmann,
Pomeron physics and QCD, Cambridge University Press (2002) 297

\bibitem{schaefer} A. Sch\"{a}fer, L. Mankiewicz, O. Nachtmann, 
Phys.Lett. B {\bf 272} (1991) 419

\bibitem{bzdak} A. Bzdak, L. Motyka, L. Szymanowski, J.R. Cudell, 
Phys.Rev. D {\bf 75} (2007) 094023

\bibitem{brodsky} S.J. Brodsky, J. Rathsman, C. Merino, 
Phys.Lett. B {\bf 461} (1999) 114

\bibitem{haegler} P. H\"{a}gler, B. Pire, L. Szymanowski, O.V. Teryaev,
Phys.Lett. B {\bf 535} (2002) 117 

\bibitem{ginzburg} I.F. Ginzburg, I.P. Ivanov, N.N. Nikolaev, 
Eur.Phys.J. C{\bf 5} (2003) 02



\end{thebibliography}
\end{document}